%% file: BCCA24.tex
\documentclass[conference]{IEEEtran}
\usepackage{cite}
\usepackage{amsmath,amssymb,amsfonts}
\usepackage{algorithmic}
\usepackage{graphicx}
\usepackage{textcomp}
\usepackage{xcolor}
\usepackage{booktabs}
\usepackage{adjustbox}
\usepackage{comment}
\usepackage{multirow}
\usepackage{tabularx}
\usepackage[]{algorithm2e}
\usepackage{enumitem}
\usepackage{times}
\usepackage{epsfig}

\usepackage{hyperref}

\def\BibTeX{{\rm B\kern-.05em{\sc i\kern-.025em b}\kern-.08em
    T\kern-.1667em\lower.7ex\hbox{E}\kern-.125emX}}

\IEEEoverridecommandlockouts    
\begin{document}

\title{Decentralized Biometric Authentication based on Fuzzy Commitments and Blockchain
\thanks{This work was partially supported by project SERICS (PE00000014) under the MUR National Recovery and Resilience Plan funded by the European Union - NextGenerationEU.}}

%\and

\author{\IEEEauthorblockN{Nibras Abo Alzahab, Giulia Rafaiani, Massimo Battaglioni, Franco Chiaraluce, Marco Baldi}
\IEEEauthorblockA{Department of Information Enginering \\ Universit\`{a} Politecnica delle Marche, Ancona (60131), Italy \\ \texttt{\{n.abo\_alzahab, g.rafaiani\}@pm.univpm.it} \\ \texttt{\{m.battaglioni, f.chiaraluce, m.baldi\}@univpm.it}}}

\maketitle

\begin{abstract}

Blockchain technology, which was introduced for supporting cryptocurrencies, today provides a decentralized infrastructure for general information storage and execution of algorithms, thus enabling the conversion of many applications and services from a centralized and intermediated model to a decentralized and disintermediated one.
In this paper we focus on biometric authentication, which is classically performed using centralized systems, and could hence benefit from decentralization.
For such a purpose, however, an inherent contradiction between biometric applications and blockchain technology must be overcome, as the former require keeping biometric features private, while blockchain is a public infrastructure.
We propose a blockchain-based biometric authentication protocol that enables decentralization and resilience while protecting the privacy, personal data, and, in particular, biometric features of users.
The protocol we propose leverages fuzzy commitment schemes to allow biometric authentication to be performed without disclosing biometric data.
We also analyze the security of the protocol we propose by considering some relevant attacks. %with practical implementation in our \textbf{GitHub} repository: \\\url{https://github.com/secomms/decentralizedbiometrics.git}
\end{abstract}


\section{Introduction}

Biometric authentication, for example based on fingerprints or facial images, has now entered into common, everyday use, and is effectively helping to mitigate the problem of managing traditional secret credentials, such as passwords.
Biometric authentication is based on the initial acquisition of a template associated with one or more of the user's biometric features and on the comparison of subsequent acquisitions of the same biometric features with the initial template \cite{jain2004introduction}.
From the biometric authentication paradigm itself, two fundamental and opposed requirements naturally arise: template protection and authentication portability.
In fact, on the one hand biometrics must be protected even more than common passwords \cite{jain2008biometric}, since they are both credentials and personal data. On the other hand, users would like to be able to transfer their biometric authentication to different systems and devices, without having to repeat the initial enrollment. 

%\begin{enumerate}[label=\roman*)]
%    \item Template protection: biometric characteristics, in addition to being credentials, and thus confidential, are also personal data and, as such, they must be protected even better than common passwords \cite{jain2008biometric}.
%    \item Authentication portability: after initial enrolling in one place or device, users would like to be able to port their biometric authentication to other systems and devices as well, without necessarily having to repeat the initial enrollment phase. \\
%    To do so, it is encouraged to perform some normalisation techniques as suggested in \cite{jain2005score}. However, the details of this are out of the scoop of this paper.
%\end{enumerate}

A commonly adopted solution to enable biometric authentication on different systems without having to repeat the initial enrollment step is to rely on a single device performing biometric authentication (e.g., a smartphone with face, fingerprint, or iris recognition), associated with a \textit{single sign-on} system (e.g. operated by some identity provider) that allows successful authentication, once executed, to be transferred across various services.
Although this approach is very effective and widely used today, it relies on a single device able to perform biometric authentication, and this is a limitation in terms of the security and scalability of the system. In fact, individual users may be equipped with heterogeneous devices, resulting in heterogeneous authentication systems and performances. Indeed, in this case, some normalization techniques should be considered  \cite{jain2005score}. In addition, each individual device represents a single point of failure and may be subject to theft or loss.

Blockchain and distributed ledger technology (DLT) represent an important innovation and provide a decentralized digital infrastructure characterized by the absence of single points of failure. Originally introduced as a support for decentralized electronic cash \cite{nakamoto2008bitcoin}, the blockchain technology has been soon used as a general-purpose infrastructure for data storage and certification, as well as for decentralized execution of algorithms known as smart contracts \cite{szabo1997formalizing}. Blockchain and DLT have already been applied for disintermediating many applications and services that are classically intermediated, i.e., operated by some central authority. For example, we can mention storage networks, supply chain management, voting systems, social networks, and so on. The initial model of a blockchain, in which all peers are equivalent and all data are public, has been extended into a variety of other models that fall under the more general umbrella of DLT, thus enabling users' differentiation and access control over data.
Based on the rights to access the ledger with reading and/or writing permissions, blockchains can be divided in different types. In particular, blockchains are defined \textit{public} if everyone can read the information contained in them, or \textit{private} if only some authorized users can read it. At the same time, blockchains can be \textit{permissionless} when everybody can join them and write new data, or \textit{permissioned} if a user must be authorized to join the network and to write data in the ledger according to some privileges.

In this paper, we propose a solution to implement a decentralized and disintermediated biometric authentication system that leverages the blockchain technology.
The main advantage over classic, centralized biometric authentication systems is that of allowing each user, after an initial enrollment phase, to be authenticated from any device participating in the network, rather than being limited to the one they initially registered with.
The main challenge to achieve this goal is dealing with the public nature of the data stored in the blockchain, which is not compatible with the sensitive nature of biometric data.

To overcome such an issue, we leverage fuzzy commitment schemes (FCSs), first introduced by Juels and Wattenberg in \cite{juels1999fuzzy}. Briefly, a FCS is a protocol based on cryptographic primitives that allows two parties to commit to a value while adding a degree of fuzziness. Differently from a conventional bit commitment scheme, indeed, in the FCS the commitment can be opened using a witness that is not identical to the original encrypting one, but it is near according to some defined metric. For this reason, FCS is suitable for biometric authentication systems, where the input biometric data slightly differ for every new acquisition. 
Due to its features, FCS allows performing biometric authentication in the encrypted domain. This is achieved by combining cryptographic primitives with error-correcting codes (ECCs) to deal with the variability of biometric features and allowing a biometric acquisition to be compared with the original template without either storing or transmitting them in clear \cite{chauhan2019improved}.

The main novelty of this work is to propose for the first time the joint use of FCSs and blockchain, along with smart contracts, to design a protocol able to achieve decentralized biometric authentication, while complying with the strict privacy requirements arising from the sensitiveness of biometrics.

The paper is organized as follows: in Section \ref{related} we describe the related works and the contribution of our paper. In Section \ref{sec:Prelis} we recall some preliminaries on  FCSs and ECCs. In Section \ref{sec:ProblemStatment} we describe each stage of the protocol we propose, for which the implementation and the experimental results are shown in Section \ref{sec:Functions}. Moreover, a security analysis is provided in Section \ref{sec:security}. Finally, in Section \ref{conclusion} we give some conclusive remarks.

\section{Related works}
\label{related}

Let us describe the related works currently existing in the state of the art and show how our approach improves upon them. 
Few studies to date have looked at the integration of biometrics and blockchain.
One of them is \cite{delgado2019blockchain},  which covers costs, privacy, processing capabilities, scalability, and security of using blockchain technology for biometrics. Biometric templates protection is also discussed. The authors show that the modifications to existing biometric systems can be kept to a minimum, allowing the use of standard biometric techniques and algorithms.
The proposed protocol does not need complex smart contracts, simplifying development and lowering execution costs.
%However, biometric authentication is performed off-chain, which still represents a possible single point of failure.\todo{Check} \todo{I think the information provided is accurate and indeed they performed the authentication off-chain} \textcolor{purple}{I think that this statement is a bit too strong, especially since we perform authentication off-chain too!} \textcolor{cyan}{If so, we should highlight the difference between our work and theirs}
%\textcolor{purple}{This is a bit confusing for me. Maybe it can be written in a different way, since for me here we are saying ``storing hashes on blockchain is unsuitable", but that's exactly what we are doing.} \textcolor{blue}{Here it says about direct hashing. But in our work, we are hashing the commitment, not the biometric}
A follow-up of the analysis in \cite{delgado2019blockchain} is presented in \cite{delgado2020blockchain}, discussing the main storage schemes for public blockchains (Ethereum) and implementing a smart contract to estimate storage costs. The analysis shows that simple schemes like direct storage of biometric templates on-chain or direct biometric data hashing are unsuitable for real-world biometric systems.
Instead, when more advanced data structures such as Merkle trees are used, the storage cost is fixed regardless of the total data volume, also reducing the execution time. Reading operations (retrieving) of templates are typically free and very fast to execute because they are performed locally.
Furthermore, the performance and cost factors of i) off-chain and on-chain biometric matching and ii) storing the biometric templates with or without template protection are studied, both qualitatively and quantitatively. Moreover, it is shown that adequate template protection could maintain and even improve biometric matching accuracy compared to unprotected templates (as recommended by best practices in privacy-preserving biometrics and related legislation such as the European Union General Data Protection Regulation (GDPR)). %For simple classifiers based on Hamming distance, on-chain matching has been shown to be feasible and cost-effective.
%\textcolor{purple}{Also here I see an issue. We point out how on-chain matching is good, but then we perform it off-chain. So maybe we can remove this sentence?} \textcolor{blue}{I do agree with you, I have commented it.} 
%\todo{This work seems `dangerous', we shall identify and clearly state some advantages of our approach}
%The proposed framework outperforms the benchmark by 55.56\% in securing biometric templates during data transmission between the enrollment device and the node database.
The authors in \cite{lee2021}, instead, divide the biometrics templates in different fragments, which are stored off-chain by different nodes. The blockchain is then used to manage these fragments.
Paéz et al. proposed a blockchain-based architecture for biometric electronic identification documents \cite{paez2020}. In their approach, a private blockchain is used to register and verify all the transactions made by users who are identified off-chain. Therefore, biometrics is not strictly part of the blockchain architecture.
The authors of \cite{nandakumar2017} proposed a biometric token for user verification. The security of the approach is based on the fact that the token can be used only once.
Bao and You presented a two-factor authentication method based on blockchain and fuzzy extractors \cite{bao2021}; they require the user to have also a password and the chosen blockchain is Fabric, thus a permissioned infrastructure.
Another model exploiting biometrics and blockchain was proposed in  \cite{mohsin2019based}; it considers an access point (patient enrollment device) and a node database for patients' authentication. A hybrid biometric pattern model based on a merge algorithm that combines radio frequency identification and finger vein biometric features is proposed to increase pattern structure randomization and security. Following that, a hybrid pattern model using a combination of encryption, blockchain, and steganography techniques is introduced. 
The recent work \cite{kaur2023blockchain} considers the integration of biometrics and blockchain, considering 3D face and ear biometrics.
In that work, however, the classical approach of blockchain-based data notarization is followed, by writing into the blockchain a simple hash digest of the biometric features, which are stored off-chain.
This requires keeping biometric features in a separate (and likely centralized) repository, which must be accessible each time an authentication is performed. 
In contrast, our approach allows biometric authentication to be performed by leveraging only the data stored in blockchain.
%\textcolor{purple}{Nibras, please add a row in the Table I and fill it accordingly}
%\todo{A clearer explanation of this work and its differences with ours is needed} \todo{I have re-read the paper and summarized the main drawbacks of the proposed methods of their work. I have added a paragraph in the contribution section stating that our scheme can overcome all of it.}\\
The main proposals regarding storage  of biometrics could be summarized as follows \cite{delgado2020blockchain}:
\begin{enumerate}
    \item full on-chain storage: templates are stored in blockchain; %very costly
    \item data hashing: the templates are stored off-chain and their hashes are stored on-chain, to provide integrity guarantees (this is also the case of \cite{mohsin2019based}); %lack of availability due to the local storage \textcolor{purple}{I can't understand the meaning of this point. We perform kind of data hashing too} \textcolor{blue}{What was meant here is that this approach can lead to a lack of availability of the biometric data, as the original data is stored locally and must be retrieved and compared to the hash, In our case, we are not storing the hash of the biometric. Instead, we are storing the commitment. Therefore, there is no local biometric in our case. However, I agree with you that it is unclear; I suggest deleting it}.
    \item linked data structure: the templates are used, for example, as nodes of a Merkle tree.
    %the structure needs to be recalculated whenever a biometric is added, deleted or modified.
\end{enumerate}

Among these, full on-chain storage is obviously inefficient and costly. Data hashing requires the availability and integrity of the templates stored off-chain and their corruption or loss would compromise the effectiveness of the system. Finally, linked data structures require the recalculation of the root every time a template is added, deleted or modified. We discuss next how our approach overcomes all these issues.

%. In fact, the storage of the hash is more efficient in terms of costs. \textcolor{cyan}{MASSIMO:I think Giulia was referring to this, right? Before we said storing the hash was not convenient} \textcolor{blue}{Exactly}

%Additionally, we do not need to store anything outside the blockchain, differently to the data hashing mentioned in \cite{delgado2020blockchain} \textcolor{blue}{This point explains a bit the point 2 in biometric storage}.

Our protocol overcomes the drawbacks of performing full on-chain data storage, since for each biometric of any user we only store a hash value and the so-called offset of a FCS, whose length $N$ is the block length of the employed ECC.
The advantage over schemes based on direct data hashing is that, in our protocol, there is no need to preserve the enrolled biometric templates to check for data integrity,  owing to the employment of the FCS. All the data necessary for user authentication (which, as mentioned above, are a hash and an $N$-uple) are stored on-chain and the employed blockchain can be public or private, as further discussed in Section \ref{sec:blockchain_type}. This also reinforces the availability and security of the data.
Unlike linked data structures, we do not need to recalculate data during any update. Furthermore, the method we propose performs the matching computation off-chain, yielding a significantly less expensive procedure with respect to the on-chain computation. 
To the best of our knowledge, in the literature there are no previous works that combine together \textit{all} the three technologies we consider, i.e., biometric authentication, FCS, and blockchain, especially considering that our approach works also leveraging public blockchains.

\section{Preliminaries}
\label{sec:Prelis}

\input{FCS}

\section{Proposed model}
\label{sec:ProblemStatment}

The system we propose in instantiated through a smart contract (SC) stored and executed on the blockchain, which is responsible for maintaining and managing a list of blockchain nodes acting as enrollment centers (ECs) and authentication centers (ACs).
%by writing their blockchain addresses into an AC list managed by the SC. The SC is indeed necessary to efficiently manage the lists of the authorized ACs and ECs.
Each EC is responsible for the enrollment of new users by registering %their blockchain addresses along with 
their biometric templates (in the encrypted domain, through fuzzy hashing, as  detailed later in Algorithm \ref{alg: Enrollment Stage}), into a user list, which is also maintained and managed by the SC.
%generates an array of registered subjects (with type subject)
%The enrollment center invites subject $S(n)$ to register their biometrics data. During the registration, each subject provides his biometric data. The enrollment center extracts features, and then builds a template. Afterwards, perform fuzzy hash and generates fuzzy data to be stored.
For each enrolled user, the EC also includes in this list an ID with 
 the corresponding fuzzy data. 
 ACs can then perform biometric authentication on enrolled users. 
For this purpose, the AC acquires the user's biometric data and extracts the feature vectors on a local host. The AC then retrieves the fuzzy hash from the blockchain and locally performs computations to verify whether the acquired and the retrieved biometrics match. 

\begin{figure}[ht]
    \centering    \includegraphics[width=\columnwidth]{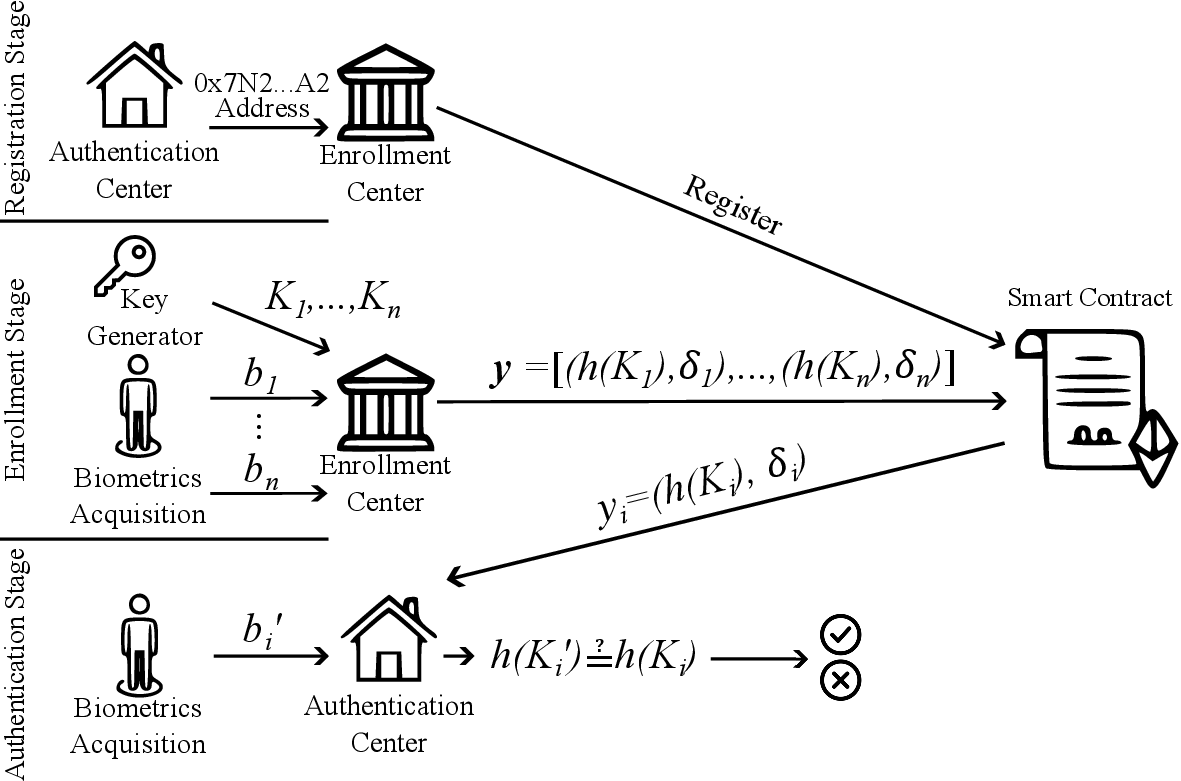}
    \caption{General scheme of the stages of the proposed system for decentralized biometric authentication of a generic user.}
    \label{fig:BlockDiagram}
\end{figure}

The entire functionality of the system, which is summarized in Fig. \ref{fig:BlockDiagram},  is divided into an initial setup and three main stages: i) registration, ii) enrollment and iii) authentication. These stages may be followed by a revocation stage.
Next, we provide a more detailed description of each stage.

\subsection{Initial setup}

%\textcolor{red}{Explain that a governmental body initializes the system by deploying the smart contract onto a public blockchain and including one or more initial ECs into the smart contract list.Also mention that government-delegated ECs are needed to ensure that ECs are not malicious and to perform initial users' identification. } \textcolor{blue}{modified completly}

%\todo{Change (next and everywhere) the private blockchain into a public one}
%The proposed model leverages a consortium blockchain, through which 
%The system is initialized by deploying a smart contract responsible for collecting and maintaining the data of participating users.
%The smart contract provides a payable function that allows new data to be written into the blockchain and some non-payable functions used to retrieve data from the blockchain.
%The payable function can be invoked by the subset of nodes constituted by the ECs, which are responsible for the enrollment of end users and of ACs. The latter are limited blockchain nodes that cannot write on the blockchain but are responsible for performing users' authentication based on data retrieved from the blockchain.

An entity that wants to instantiate the system can initiate it by deploying the smart contract onto a blockchain and by adding one or more initial trusted ECs to the smart contract list. The ECs delegated by the creator entity are initially required to ensure that newly registered ECs are not malicious and to perform user enrollment. The smart contract is in charge of collecting and maintaining the data of participating users, and it provides a payable function for writing new data into the blockchain, as well as non-payable functions for retrieving data from the blockchain. The payable function can only be used by a subset of nodes, which are made up of ECs. These ECs are in charge of enrolling end-users and ACs, whereas ACs are limited blockchain nodes that cannot invoke payable functions, but are in charge of performing user authentication based on data retrieved from the blockchain.

\subsubsection*{
Interoperability with different  blockchain types}
\label{sec:blockchain_type}
We would like to emphasize that,  with some slight adjustments, the proposed protocol is compatible with any type of blockchain. 
In fact, it is possible to use a public permissionless blockchain  since, thanks to the use of FCS, the data stored in the ledger are in the encrypted domain, ensuring there is no information leakage. This allows for maximum accessibility, reliability, and transparency of the system. However, although anyone can freely join the network in this scenario, there is a risk of malicious users impersonating an AC or an EC. This could result in a situation where a user cannot be certain that they are  authenticating through a legitimate and authorized center. To solve this issue, a public list of authorized ECs and ACs, together with their blockchain addresses, should be created and spread. This would allow users to simply consult the list and confirm that they are referring to a trusted authority. It is also worth noting that the use of public permissionless blockchain comes with some costs. Including transactions in such blockchains, as well as deploying smart contracts, indeed, need the payment of some fees through cryptocurrencies. The benefits of the employment of diffused blockchains (i.e., Ethereum) should be weighted against these non-negligible costs. 
Instead, the use of private and permissioned blockchains leads to a drastic reduction of these costs, that can be proximal to zero, while limiting full decentralization and 
accessibility. Moreover, by allowing the possibility to decide who can access the network and how, private blockchains perform better from a security point of view, thus resulting in a system in which all the nodes are trusted. 
Therefore, according to the considered context and needs, it is possible to implement the proposed protocol leveraging different kinds of blockchain.

In the rest of the paper, we will consider a consortium blockchain, i.e, a public blockchain where only some authorities have writing rights. However, we highlight once again that different types of blockchains can be used without compromising the functionalities of our protocol.

\subsection{Registration stage}

\begin{figure}[htb]
    \centering
    \includegraphics[width=0.8\columnwidth]{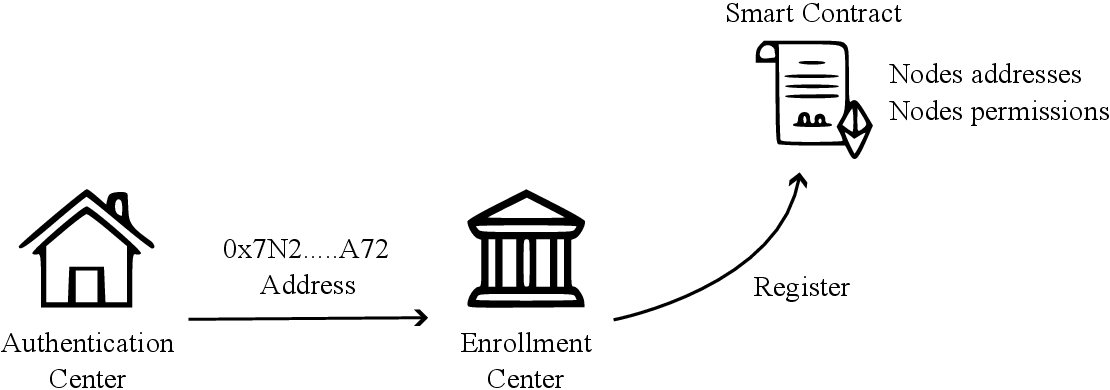}
    \caption{Registration Stage.}
    \label{fig:Regestration}
\end{figure}

The registration stage allows new ACs, and new ECs as a special case, to be incorporated into the system.   %\textcolor{blue}{with the permission of the EC, which invoked the smart contract.} \textcolor{cyan}{But only governmental-allowed ECs can give these permissions? Or any EC?}
In fact, by invoking the smart contract, an existing EC is responsible for storing onto the blockchain the addresses and the data of new ACs. The registration steps are described in Fig. \ref{fig:Regestration}. %\textcolor{cyan}{I would remove the ``Registration stage'' on the top left of the figure, it is redundant.}
%\todo{There are typos in this figure too}\todo{Modified and reuploaded}
Each newly registered AC is provided with read-only permissions on the consortium blockchain.
As a special case of registration, after a voting session where all the ECs cast their votes, the registered node can be provided with writing permissions. In this case, the AC is elevated to the same hierarchical level of the EC, thus becoming a fully-fledged EC. %\textcolor{cyan}{maybe I am missing something but Marco's comment made me think that ECs needed some kind of governmental permission to be included. In this description it seems that it is sufficient to have the permissions by other ECs. This should be clarified.} \textcolor{blue}{I have added a sentence to explain this, I hope it is more clear now}

\begin{comment}
\begin{algorithm}[ht]
\fbox{
\begin{minipage}{0.9\columnwidth}
\DontPrintSemicolon
\textbf{BEGIN} \;
\SetKw{Kwstructure}{structure}
\SetKwProg{Str}{Structure}{:}{}
\SetKwFunction{Node}{Node}
\Str{\Node}{
\textit{Public} \quad ID \;
\textit{Public} \quad Name \;
\textit{Private} \quad Adress \;
\textit{Private} \quad IsAuthentication \;
\textit{Private} \quad IsEnrollment \;
}
\;

Center\_1  \quad $\gets$  \textbf{NEW} \quad \Node \;
Center\_1.$ID \quad \gets 1$\;
Center\_1.$Name \quad \gets$ \textbf{INPUT}\; 
Center\_1.$Address \quad \gets$ \textbf{INPUT}\; 
Center\_1.$IsAuthentication \quad \gets$ True\; 
\textcolor{gray}{\textit{\# If it will be an Authentication Node}}\;
Center\_1.$IsEnrollment \quad \gets$ True\; 
\textcolor{gray}{\textit{\# If it will be a new Enrollment Node}}\;

\textbf{WRITE}.To\_Blockchain(Center\_1)\;

\textbf{END}\;
 \end{minipage}
 }
 \caption{Algorithm for Registration Stage}
 \label{alg:Regestration Stage}
\end{algorithm}
\end{comment}

\subsection{Enrollment stage}

The goal of this stage is to enroll users in the system. Enrollment of the $j$-th user based on the FCS can be performed by any EC, and starts by generating a random key $K$ (the witness, according to Algorithm \ref{algoFCS}) when a new user requires to be enrolled into the system. 
The corresponding codeword $c$ is then XORed with the subject's biometric information $x$, which represents the features extracted from the acquisition of some biometric data $b$, to generate the offset $\delta$. At the end of the enrollment stage, the offset $\delta$ along with the hash of the key $h(K)$ are stored onto the blockchain. This process can be also applied to $n$ different biometrics for every user, so that the authentication protocol is not limited to just one specific type of biometrics. 
%each biometric feature associated to the user ID. 
Note that, owing to the features of the cryptographic functions at the basis of the protocol, as briefly discussed in Section \ref{sec:security} and more extensively in \cite{juels1999fuzzy}, it is unfeasible for an attacker to extract user biometric information from the data stored in the smart contract (and, thus, in the blockchain).

\begin{figure}[htb]
    \centering
    \includegraphics[width=\columnwidth]{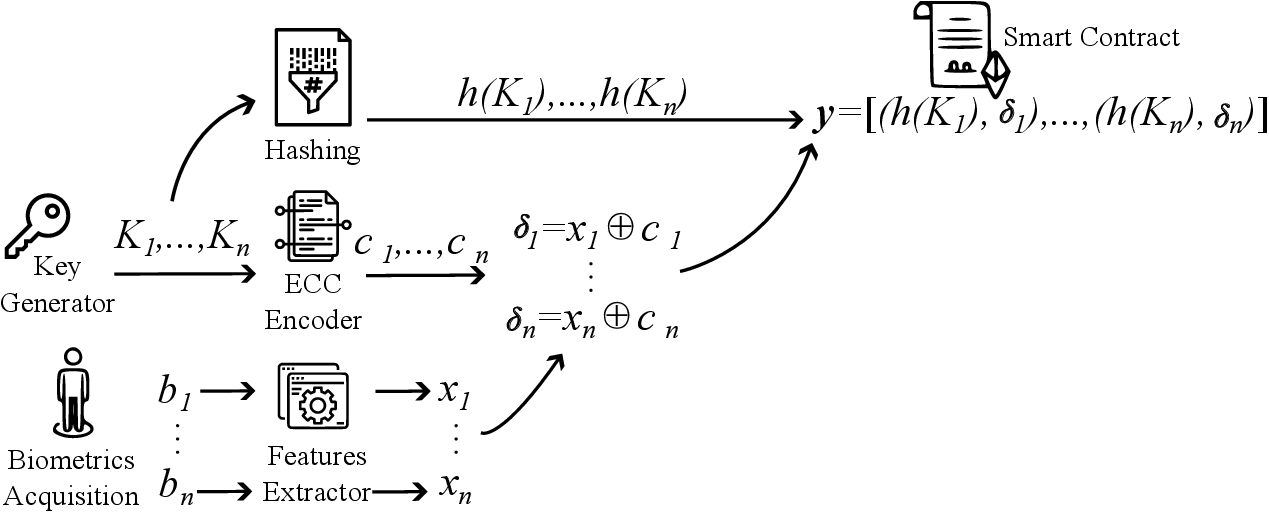}
    \caption{Enrollment Stage of a generic user.}
    \label{fig:Enrollment}
\end{figure}

The enrollment stage is depicted in Fig. \ref{fig:Enrollment} and described in detail in Algorithm \ref{alg: Enrollment Stage}.   It is worth noting that all the functions can be performed off-chain, except for the writing function, which instead requires interaction with the smart contract.

\begin{algorithm}[htb]
\centering
\resizebox{0.4\textwidth}{!}{
\fbox{
\begin{minipage}{0.4\textwidth}
%\textbf{BEGIN} \;
\DontPrintSemicolon
%\SetKw{Kwstructure}{structure}
\SetKwFunction{Subject}{Subject}
\SetKwProg{Str}{Structure}{:}{}
\SetKwProg{Fn}{Function}{:}{}
\SetKwFunction{KeyGenerator}{KeyGenerator}
\SetKwFunction{ECCEncoder}{ECCEncoder}
\SetKwFunction{FeaturesExtractor}{FeaturesExtractor}
\SetKwFunction{HashFunction}{Hash}
\SetKwFunction{Open}{Open}
\SetKwFunction{Commit}{CommitFSC}

\Str{\Subject}{
Public $\mathrm{ID}$ \;
Private $\mathrm{HK}$ \;
Private $\mathrm{Delta}$ \;
}\vspace{2mm}
%\Fn{\KeyGenerator{}}{
%        $K$ $\gets$ Generate Random Key\;
%        \KwRet\ $K$\;
%}  \;
%\Fn{\ECCEncoder{$K$}}{
%        $C$ $\gets$ Encode $K$\;
%        \KwRet\ $C$
%} \;

%\Fn{\FeaturesExtractor{$b$}}{
%        $x$ $\gets$ Extract Features\;
%        \KwRet $x$
%}  \;

%\textbf{ARRAY} $b[n]$\;
%\textbf{ARRAY} $x[n]$\;
%\textbf{ARRAY} $\delta[n]$\;
%\textbf{ARRAY} $h_k[n]$\;
%\; 

%\For{$i\gets 1 \hspace{1mm}\KwTo\hspace{1mm} n$}{
%%$K[i]$ $\gets$  \KeyGenerator{}\;
%$K[i]$ $\gets$  Generate Random Key\;
%%$C[i]$ $\gets$  \ECCEncoder{$K[i]$}\;
%$C[i]$ $\gets$  \text{Encode ($K[i]$)}\;
%$h[i]$ $\gets$ \HashFunction{$K[i]$}\;
%$b[i]$ $\gets$ \textbf{INPUT} \quad \textcolor{gray}%{\textit{$\#$ Acquire the $i$-th biometric}}\;
%%$x[i]$ $\gets$ \FeaturesExtractor{$b[i]$}\;
%$x[i]$ $\gets$ Extract Features ($b[i]$)\;
%$\delta[i]$ $\gets$ $x[i] \oplus C[i]$\;
%}    

\For{$i\gets 1 \hspace{1mm}\KwTo\hspace{1mm} n$}{
$b[i]$ $\gets$ \textbf{INPUT} \quad \textcolor{gray}{\textit{$\#$ Acquire the $i$-th biometric}}\;
$x[i]$ $\gets$ \FeaturesExtractor{$b[i]$}\;
$(h[i],\delta[i])$ $\gets$ \Commit{$x[i]$}\;
}

\vspace{2mm}
$\mathrm{Subject_j}$ $\gets$ \textbf{NEW} \Subject\;
$\mathrm{Subject_j}$.$\mathrm{ID} \gets j$\;
$\mathrm{Subject_j}$.$\mathrm{HK} \gets$ $h$\;
$\mathrm{Subject_j}$.$\mathrm{Delta} \gets \delta$\;

%\textbf{WRITE}.To\_Blockchain($\mathrm{Subject_1}$)\;
\vspace{2mm}
\textbf{WRITE} ($\mathrm{Subject_j}$)\;

%\textbf{END}\;
\end{minipage}
}
}
 \vspace{1em}
 \caption{Pseudocode of the Enrollment Stage.}
 \label{alg: Enrollment Stage}
\end{algorithm}

\subsection{Authentication stage}

\begin{figure}[ht]
    \centering
    \includegraphics[width=0.95\columnwidth]{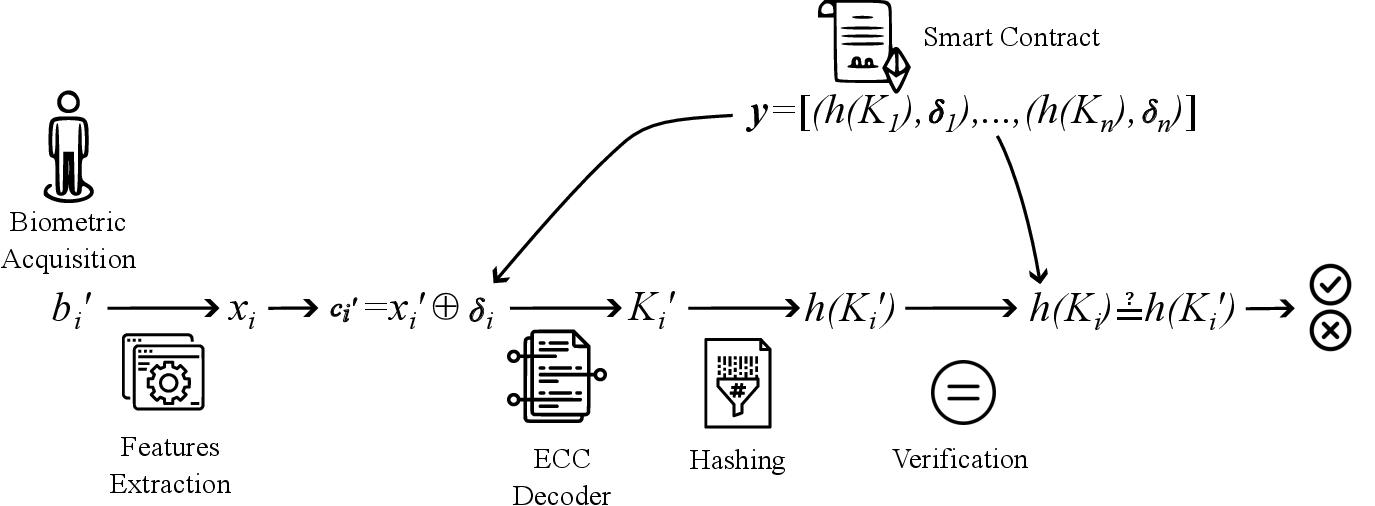}
    \caption{Authentication Stage for a generic user.}
    \label{fig:Authentication}
\end{figure}

This stage allows any  enrolled user to be authenticated by any AC. The user being authenticated communicates their ID to the AC, and some new biometric feature  $x'$ is acquired.
The AC, thanks to the ID, retrieves the corresponding values of $\delta$ and $h(K)$ from the smart contract. 
According to the FCS, the biometric feature $x'$ is then XORed with the offset $\delta$ to generate $c'$, which is given as input to the ECC decoder that, in its turn, returns an estimated codeword $\widetilde{c}$. The corresponding estimated key $K'$ can be easily obtained from $\widetilde{c}$ (for example, by leveraging systematic encoding). 
If the new biometric acquisition is close to the original one (according to some defined metric), then the ECC decoder is likely to retrieve the original codeword $c$ as $c'$, implying $K' = K$. % Then, $C$ can be demapped into the correct key $K$.
By comparing the digest of $K'$ with the value of $h(K)$ retrieved from the blockchain, the AC can hence decide whether the user authentication is successful or not.
%\textcolor{purple}{
The result of the authentication stage can be submitted to the blockchain for logging. This can be done, for example, when a private blockchain is used, so that the costs will be contained. %} \textcolor{cyan}{We agreed that this should be optional (depending on the considered type of blockchain). We will rewrite this sentence after having written the paragraph on blockchains.} \todo{Do we really need this final step?} \todo{I think yes, it can give us a track for the authentications performed. For example, a user can keep track of the times he was authenticated. He can recognize any suspicious authentication he did not do by himself. Additionally, it increases the transparency between the system and its users.}
The authentication stage is detailed in Fig. \ref{fig:Authentication}, and formalized in Algorithm \ref{alg: Authentication Stage}, where the writing and the reading functions involve the smart contract, while all the other operations are carried out off-chain.

\begin{algorithm}[ht]
\centering
\resizebox{0.4\textwidth}{!}{
\fbox{
\begin{minipage}{0.4\textwidth}
%\textbf{BEGIN} \;

\DontPrintSemicolon
\SetKwFunction{FeaturesExtractor}{FeaturesExtractor}
%  \SetKwProg{Fn}{Function}{:}{}
%  \Fn{\FeaturesExtractor{$Biometric$}}{
%        $X$ $\gets$ Do Features Extraction Procedures\;
%        \KwRet $X$
%  }
 % \;
  \SetKwFunction{ECCDecoder}{ECCDecoder}
  \SetKwFunction{Open}{OpenFCS}  
  \SetKwProg{Fn}{Function}{:}{}

%  \Fn{\ECCDecoder{$C$}}{
%    \textcolor{gray}{\textit{$\#$ Error Correcting Code Decoder}}\;
%        $K$ $\gets$ Decode $C$\;
%        \KwRet $K$
%  }
%  \;
%  \SetKwFunction{HashFunction}{HashFunction}

$\mathrm{ID}$ $\gets$ \textbf{INPUT}\;
$b'[i]$ $\gets$ \textbf{INPUT} \quad \textcolor{gray}{\textit{$\#$ Acquire the $i$-th biometric }}\;

$(\delta[i],h[i])$ $\gets$ 
%\textbf{READ}.from\_Blockchain($ID$)\;
\textbf{READ} ($\mathrm{ID},i$)\;
\vspace{2mm}
$x'[i]$ $\gets$ \FeaturesExtractor{$b'[i]$} \;

$\mathcal{O}[i]$ $\gets$ \Open{$\delta[i]$,$x'[i]$,$h[i]$}
%$C'[i] \gets x'[i] \oplus \delta[i]$\;
%$K'[i] \gets$ Decode ($C'[i]$)\;
%$h'[i] \gets$ \HashFunction{$K'[i]$} \;

\vspace{2mm}
\eIf{$\mathcal{O}[i]$}{
    \textbf{DO} Authenticate User \;
    %\textbf{WRITE}.To\_Blockchain(Authenticated
    \textbf{WRITE}(Authenticated)\; 
  }{
    \textbf{DO} Reject User \;
    %\textbf{WRITE}.To\_Blockchain(Not Authenticated)\;
    \textbf{WRITE}(Not Authenticated)\;
  }
%\textbf{END}\;
 \end{minipage}
}
}
\vspace{1em}
 \caption{Pseudocode of the  Authentication Stage.} %\textcolor{cyan}{should not $i$ be in some way an input to READ(ID) also? In such a wa}
 \label{alg: Authentication Stage}
\end{algorithm}

\subsection{Revocation stage}

According to GDPR and data protection best practices, the right to be forgotten from the system must be guaranteed to each user.
In the framework we propose, this can be ensured by ECs. 
In fact, any EC can invoke a payable function of the smart contract that erases the user records from the list of enrolled users.
Despite ACs will no longer be able to retrieve data concerning revoked users and thus perform their authentication, their enrolling data will still be stored within past transactions.
However, according to the fuzzy commitment paradigm, these data are stored in encrypted form, and no personal data of revoked users can be retrieved.
The system may also require to revoke the privileges of some AC or even EC.
In both cases, a voting mechanism involving the nodes present in the network should be implemented to make a majority decision and update the list of ACs or ECs accordingly. This will be further explored in future work.

\section{Implementation and Experiments}
\label{sec:Functions}
The proposed model, described in Section \ref{sec:ProblemStatment}, was implemented as a decentralized application (dApp). The considered smart contract was developed using  \texttt{Solidity v0.8.21} and deployed using \texttt{Truffle v5.11.5}, with testing conducted using \texttt{Node v16.20.2} and \texttt{Web3.js v1.10.0}. Additionally, smart contract performance was also evaluated using \texttt{Python v3.10.12} and \texttt{web3[tester]  v6.15.1}.
The source code for implementing and testing the system is publicly available%please refer to our \textbf{GitHub} repository 
\footnote{\url{https://github.com/secomms/decentralizedbiometrics.git}}.
%We list a brief description of the 
The smart contract provides the following list of functions:
\begin{enumerate}
    \item \textbf{Functions for Subjects}.
    The contract provides several functions to interact with subjects:
      \begin{itemize}
        \item \texttt{setSubjects}: allows an EC to set subject commitment.
        \item \texttt{getSubjects}: permits an authorized AC to retrieve subject information.
        \item \texttt{updateSubjects}: allows an EC to change the commitment of an existing subject.
        \item \texttt{delSubjects}: allows an EC to delete subject information.
      \end{itemize}
    \item \textbf{Functions for Nodes} (ACs and ECs). Similar to subjects, functions for managing ACs and ECs are provided:
      \begin{itemize}
        \item \texttt{setNodes}: allows an EC to set information for ACs.
        \item \texttt{getNodes}: enables an EC to retrieve AC information.
        \item \texttt{updateNodes}: enables an EC to change the status of an AC to have EC's privileges. 
        \item \texttt{delNodes}: allows an EC to delete AC information.
        \end{itemize} 
\end{enumerate}

%\section{Experimental Results}
%\label{results}
%After completing up the theoretical presentation of the proposed system, we explore its empirical results, concentrating on the scalability, efficacy, and efficiency of smart contracts in a blockchain context.

Then, we extensively tested the proposed smart contract's stability and functionality. %Next we provide a summary of the methods used to validate the proposed approach.
% \todo{did we anticipate that we implemented the smart contract in an actual software? We should also add the link to the github repo where we mention it for the first time.} 
We assessed different stages of the smart contract lifecycle, including deployment, registration, enrollment, authentication, revocation, node status updates, and security protections against unauthorized access, using a set of well designed tests. Furthermore, we examined gas consumption patterns for transactions that modify the smart contract's state, offering a thorough cost analysis for operating functions. Afterwards, we present the computation time of each function of the proposed smart contract. 
%This experimental results enriches our analysis by quantifying the temporal efficiency of executing each function within the smart contract. 
By comparing these aspects with existing studies, we also assess the benefits and the potential scalability of the approach we propose.
%benefits our work contributes to the realm of smart contract development.

%This section intends to showcase our contributions to the field of Dapp research and development by providing a comprehensive analysis of our findings and comparing them with previously published literature.

\subsection{Gas Consumption}

\begin{figure}[htb]
    \centering
\includegraphics[width=0.9\columnwidth]{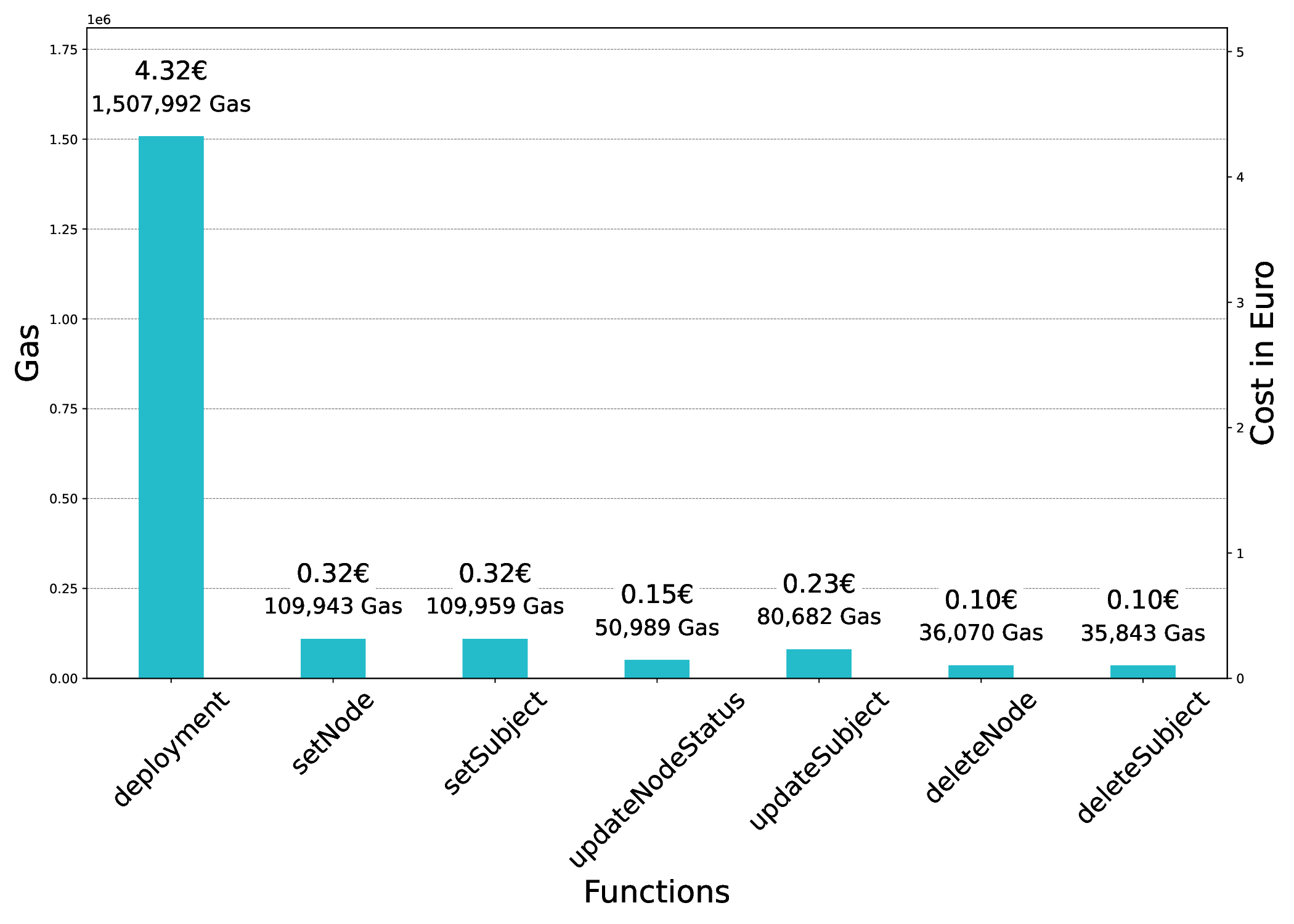}
    \caption{Gas consumption of the  smart contract deployment and payable functions execution.} %\textcolor{cyan}{Please correct deploymen into deployment..here and in the figure below}\textcolor{green}{Done}}
    \label{fig:Cost_estimation}
\end{figure}

The study of gas consumption is essential for evaluating the scalability of the smart contract, and the whole dApp system. This is focused on the operations that alter the state of the smart contract, specifically setting, deleting, or updating functions. Operations limited to retrieving data from the smart contract, instead, do not modify the contract's state and, therefore, are not subject to gas fees. %Our analysis focuses exclusively on the smart contract's payable functions.
% \todo{Including and commenting the gas analysis is important, as we know, but these paragraphs are beyond discursive, redundant and unclear. There is a lack of explanation of the functions that appear in the figure, which instead should be explained, albeit briefly, one by one before mentioning and commenting their gas consumption.
% In addition, this sort of elegant English seems AI-generated, is misplaced, and the resulting panegyrics are moreover misleading.}
Fig. \ref{fig:Cost_estimation} summarizes the costs for deploying the proposed smart contract and its functions on Ethereum. %that we described above. %described in \ref{sec:Functions}. 
%The deployment was performed on an Ethereum testnet. 
The costs are measured in Gas units and Euros (€), based on the Ethereum value as of July 10th, 2024 (i.e., $2867.49$ € per ether). This provides a detailed insight into the economic aspects of operating smart contracts within the Ethereum environment.
% \todo{figures are unclear, contain a lot of white space that could be eliminated by making them more readable, and most importantly, contain many elements that are never introduced or commented on in the text. This creates a disconnect between the content of the figures and the rest of the rest, which should not exist.}

\begin{figure}[tb]
    \centering
    \includegraphics[width=\columnwidth]{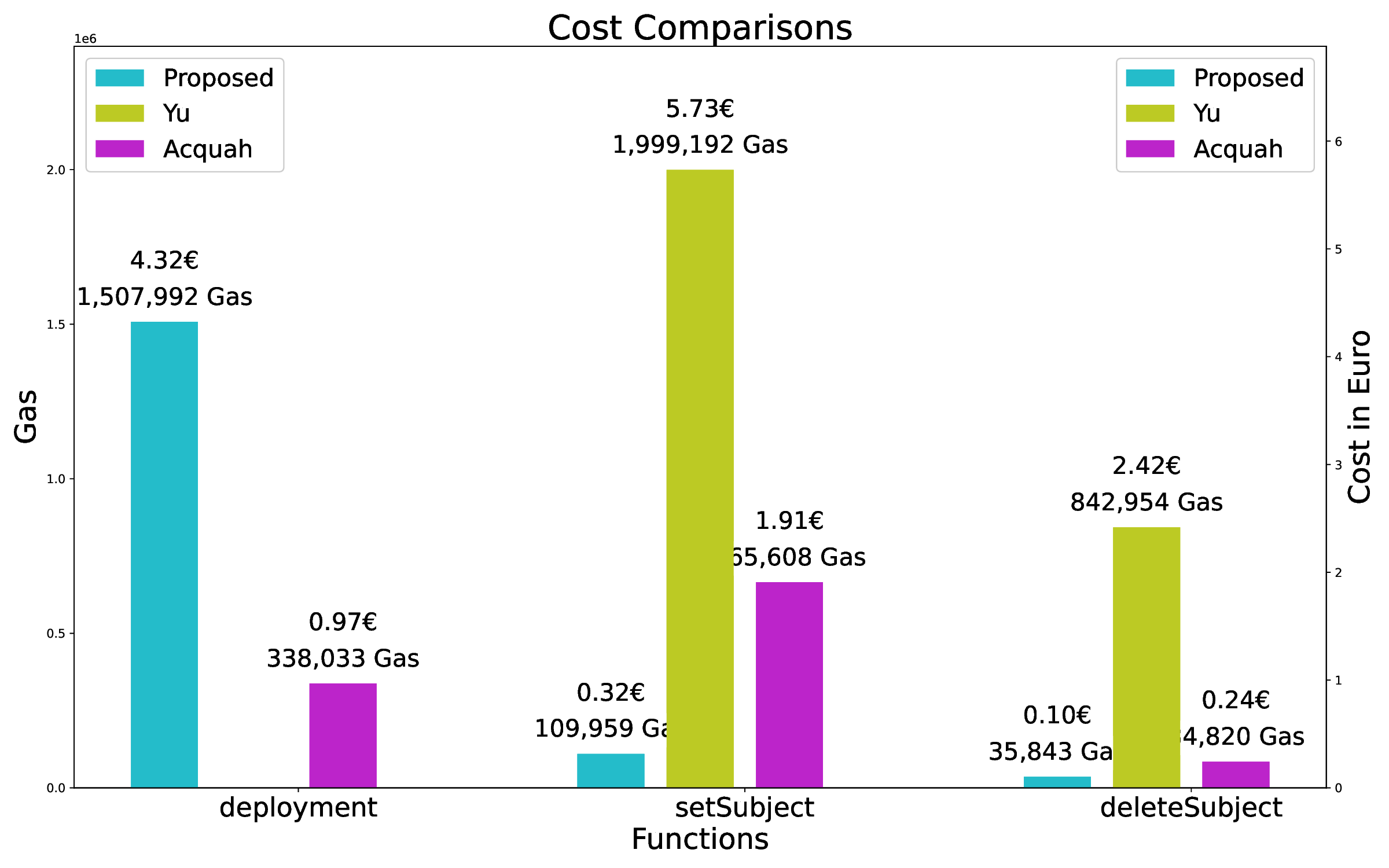}
    \caption{Gas consumption comparison with \cite{Yu2020} and \cite{Acquah2020}. Note: Ethereum price fixed at $2867.49$ € as of July 10th, 2024.}
    \label{fig:Cost_estimation_Comparison}
\end{figure}

As it is crucial to compare our work with what is already being investigated in previous works, we have compared our experimental results with the literature. In Fig. \ref{fig:Cost_estimation_Comparison} we show a comparison of gas consumption between our model and the works of \cite{Acquah2020} and \cite{Yu2020}. The comparison shows that our system has higher deployment cost compared to \cite{Acquah2020}, while \cite{Yu2020} does not illustrate cost of deployment. %\textcolor{green}{This is due to}. On the other hand, 
However, we obtain a significantly lower cost with respect to existing approaches for what concens gas consumption 
%of 50785 (0.11 €) and 35843 (0.8 €) 
for enrollment and revocation stages. %respectively. 
We highlight that the deployment function is meant to be called only once, while the other functions will be repeatedly used within the system. This shows a high potential for system scalability in terms of cost.

\subsection{Computation time}
\begin{figure}[tb]
    \centering
    \includegraphics[width=0.9\columnwidth]{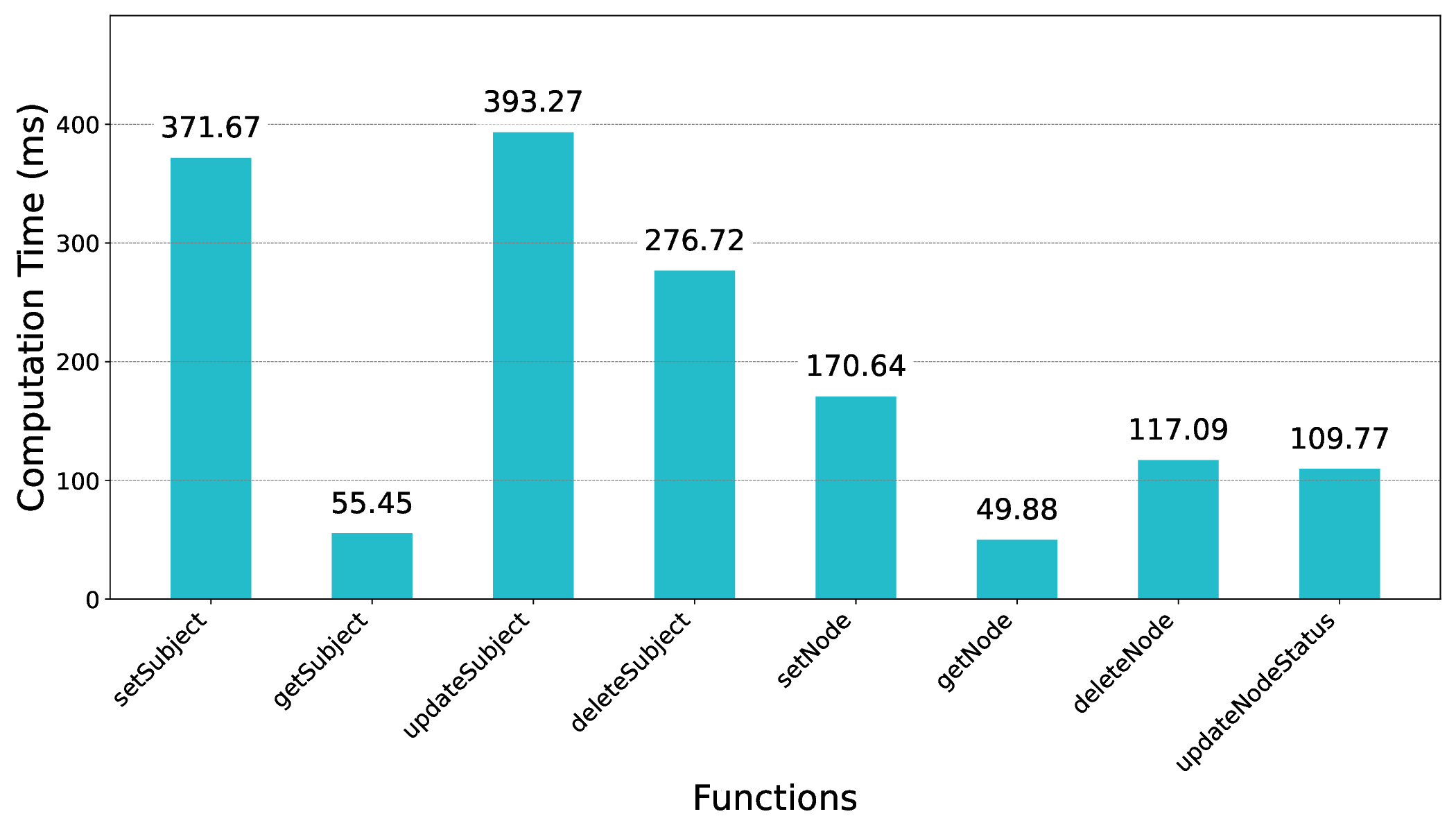}
    \caption{Transaction time (ms) for each function of the smart contract.}
    \label{fig:Computation_time}
\end{figure}

\begin{figure}[tb]
    \centering
    \includegraphics[width=0.9\columnwidth]{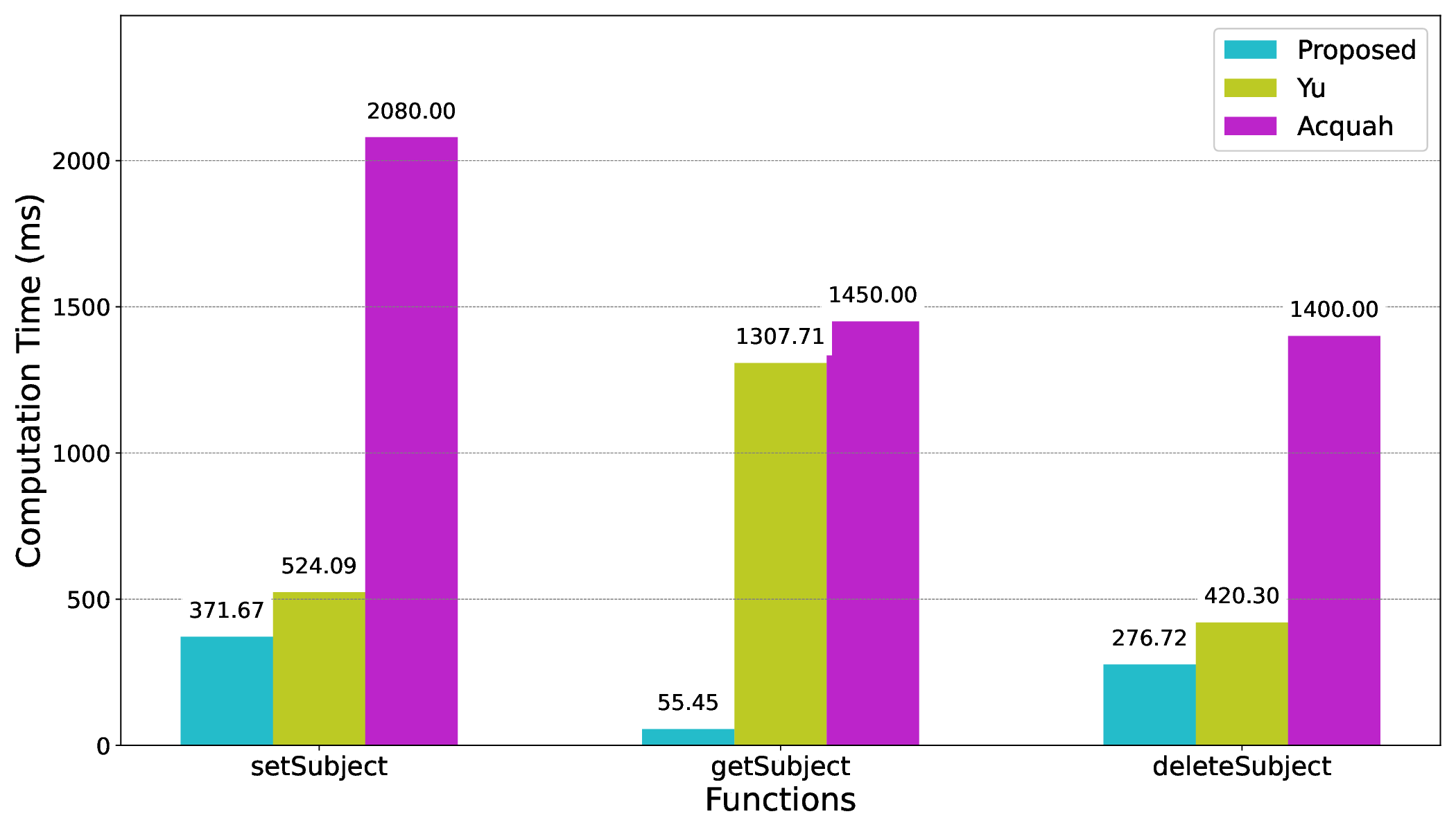}
    \caption{Transaction time comparison with \cite{Yu2020} and \cite{Acquah2020} for functions of the smart contract (ms).}
    \label{fig:Computation_time_Comparison}
\end{figure}

We then calculate the computation times for executing the proposed smart contract functions.
%of computation times for smart contract functions emerges as a cornerstone for optimizing blockchain technology's efficiency, especially in the realms of blockchain and biometrics.
Our experimental findings, illustrated in Fig. \ref{fig:Computation_time}, reveal significantly reduced computation times in comparison with \cite{Acquah2020} and \cite{Yu2020}.
This direct comparison, as demonstrated in Fig. \ref{fig:Computation_time_Comparison}, not only highlights our framework's enhanced efficiency, but also serves as a benchmark, aiding in resource management through accurate cost estimation, facilitating scalability insights, and ensuring quality assurance. Moreover, understanding computation times aids in identifying potential security vulnerabilities, optimizing for energy efficiency, and ultimately contributing to a more sustainable blockchain infrastructure. By analyzing these computation times, we offer a clearer understanding of the impact of smart contract operations on the overall system performance. %, setting a precedent for future research to build upon. 
In addition, some attack vectors on smart contracts, such as denial-of-service attacks, exploit functions that require significant computation time. Identifying and optimizing long-running functions can mitigate these security risks; this topic will also be discussed in the next section. %Consequently, our findings advocate for the critical role of computation time analysis in advancing smart contract technology, presenting implications that extend beyond mere performance metrics to encompass security, scalability, and sustainability within blockchain ecosystems.

\section{Soundness and Security}
\label{sec:security}

In this section, we focus on the analysis of the strength of the proposed approach. First, in Section \ref{soundness}, we focus on the soundness testing of the different stages involved in the protocol. Then, in Section \ref{security_analysis} we analyze the security of our approach. 
%We highlight that the code used for testing the proposed protocol is made available\footnote{Repository: \url{https://github.com/secomms/decentralizedbiometrics}}.

\subsection{Soundness testing}
\label{soundness}
To assure functionality and robustness, we used a test-oriented design to evaluate the smart contract's lifespan from deployment to operational conclusion.
For this purpose, we used simulated biometrics data.
%, emphasising its extensive security frameworks. Future biometrics will employ real data, but we tested with fake data.

% \todo{In what follows, rather than repeating the meaning of the different stages (which is already stated before) I would suggest focusing on tested guarantees and/or attack defenses.}
% \todo{Maybe even clearer, we could add a table that, for each stage, clarifies which guarantees/defenses have been tested}

\subsubsection{Deployment and Initialization} After the deployment, the smart contract is thoroughly tested to ensure it can be correctly initialized, indicating that the system is ready for usage in real-world scenarios. A trusted entity handles the initial deployment, guaranteeing early management and monitoring from a reliable source.
\subsubsection{Registration Stage} This stage evaluates the contract's capacity to add new nodes to the network. %, like post offices and hospitals. 
These nodes, which make subject enrollment and authentication easier, are essential to the system's functionality. The successful registration demonstrates the event-driven nature of smart contracts for real-time monitoring, which generates events verifying the node's inclusion.
\subsubsection{Enrollment Stage} The core functionality of enrolling new biometric identities is tested here. The process involves setting subjects with unique identifiers and biometric data, emphasizing the system's capacity to handle sensitive information securely. The events emitted upon successful enrollment serve as a proof of the contract's ability to manage biometric data efficiently.
\subsubsection{Authentication Stage} The principles of enrolling new biometric IDs are covered in this test. The process involves giving subjects unique identities and biometric data, demonstrating the system's capacity to handle sensitive data securely. %Pseudo data were utilized for testing; real biometric data would be used in subsequent editions. 
% The events that are generated following a successful enrollment serve as proof of how well the contract manages biometric data. %(Post Office, for example) and authorised organisations (white-hat hackers, for example)  draw attention to the contract's strong access control procedures.

\subsubsection{Revocation Stage} This test reinforces the security safeguards in the contract by guaranteeing that data about subjects can only be deleted by authorized entities.
\subsubsection{Node Status Update} The dynamic nature of node responsibilities within the system, which enables updating a node's status to reflect its present operational capabilities, is the main focus of this test. This flexibility depends on the system's capacity to scale and react to shifting operational requirements. 

\subsection{Security analysis}
\label{security_analysis}

%with probability significantly larger than $\frac{1}{2}$\todo{Maybe $1/2$ is referred to bit commitments schemes?}

% \todo{We had already written this part, and I believe that it is very important. But in my opinion it should be better linked with the new part "Functionality testing", which in some way also deals with security. I would put them closer together and with more connection between one each other.}
The security features of the proposed protocol are inherited from the involved technologies. Regarding the FCS, its security is based on the \textit{concealing} and \textit{binding} properties. In particular, for the concealing property, it is unfeasible for an attacker to guess the committed value without knowing the encryption key. Instead, for the (strong) binding property, it is unfeasible to decommit a value with two witnesses that are not near, i.e., are truly distant \cite{juels1999fuzzy}. Note that the security of the FCS is strictly related to the security of the underlying hash function. When applied to the proposed scheme, the concealing property ensures that an attacker, knowing only the commitments stored in the smart contract, cannot discern any information about the users' biometric features. As for the binding property, it guarantees that it is unfeasible for two different users having unique biometric attributes to decommit the same value. In other words, it is unfeasible for a user to impersonate another.

\subsubsection{Security Against Unauthorized Access}  The system's defence mechanisms against a range of unauthorized operations, such as the insertion, update, and deletion of subjects and nodes by unauthorized accounts, are validated by a set of tests performed during the testing phase. % (12 tests found in our \textbf{GitHub} repository \url{https://github.com/secomms/decentralizedbiometrics.git}).
The extensive security procedures outlined in the contract are validated by these tests, which guarantee that critical operations can only be carried out by authorized parties.

\subsubsection{CIA triad}
In the following, we discuss the so-called CIA (confidentiality, integrity, and availability) triad \cite{ham2021toward}, which is crucial for information system security, ensuring data accessibility, accuracy, and trust.

%\subsection{Confidentiality}
% \vspace{0.2em}
\paragraph{Confidentiality}
Confidentiality safeguards are intended to protect private information from unauthorized access.
Such a requirement does not comply with the public nature of the blockchain, where data are readable by anyone.
However, our proposal relies on the FCS for the protection of users' biometric data.
In the FCS, a witness is necessary to open a committed template; this is represented by the user owning the corresponding biometrics.
Therefore, when the FCS is employed, only original users can retrieve their own biometric data from the data stored in the blockchain, thus preserving confidentiality.

%\subsection{Integrity}
% \vspace{0.2em}
\paragraph{Integrity}
Integrity requires ensuring data consistency, accuracy, and trustworthiness throughout their lifecycle. Data must not be changed in transit, and precautions must be taken to prevent unauthorized people from altering data \cite{ham2021toward}.
Blockchain technology relies on the use of consensus algorithms \cite{zheng2017overview}, requiring a certain number of nodes in the network to agree on the validity of a transaction before it is added to the ledger. 
Moreover, once a transaction is stored in the blockchain, it cannot be altered. 
This helps prevent tampering and makes the proposed framework fulfil the integrity criterion. 
Concerning the smart contract lists, collecting the enabled ECs, ACs and final users, their integrity relies on the assumption that there are no malicious ECs, which might otherwise write fraudulent data or alter existing data.
Such an assumption may seem somewhat restrictive, but in reality, the absence of malicious enrollment centres is a common requirement for any digital identity system.

%\subsection{Availability}

% \vspace{0.2em}
\paragraph{Availability}
In our model, data availability benefits from the advantages of the decentralized and redundant nature of blockchains \cite{prowse2017beyond}.
In fact, each full node in the blockchain maintains a copy of the distributed ledger. 
This provides the system with resistance to single node failures and, thus, to the absence of single points of failure, ensuring high data availability.

\subsubsection{Attacks against the blockchain}

Three main types of attacks can be mounted against blockchains, namely, fork attacks, network attacks, and application attacks \cite{mastilak2022secure}.
Their possible impact on the proposed model is discussed next.

\paragraph{Fork attacks}

In this type of attack, the attacker attempts to replace the most credible chain by initiating a substitute chain in order to gain some benefit \cite{mastilak2022secure}. 
This kind of attack is prevented by making use of  to a robust blockchain infrastructure, like the Ethereum public blockchain. If a consortium blockchain is instead used to reduce operating costs, consortium members must be trusted authorities in order to avoid these kinds of attacks.

\paragraph{Network attacks}

The nodes in the decentralized network are distributed geographically. This type of attack attempts to disconnect a node or a group of nodes from the rest of the network. This could include a few attack scenarios \cite{mastilak2022secure}.
Also in this case, adopting a robust public blockchain infrastructure like Ethereum, or a consortium blockchain with trusted mining nodes, suffices to avoid these attacks.

\paragraph{Application attacks}

The smart contract underlying the entire system can also be subject to attacks.
For example, an overflow attack could cause program execution to have an undesirable outcome. 
A cautious design of the smart contract and the use of standardized data structures can prevent this type of attacks.

\subsubsection{Attacks against biometric authentication}

A further type of attacks targets the biometric information itself, aiming at stealing secret credentials and impersonating enrolled users. Some typical attacks against biometric authentication, which are relevant to our protocol, are discussed next.

\paragraph {Impersonation attacks} 

These attacks do not depend on the digital infrastructure, and substantially depend on the ease for an attacker to gather copies of the victim's biometric characteristics.
The proposed model is resilient to these attacks owing to two of its inherent characteristics: i) protection of biometric features through the FCS and ii) suitability for multi-model biometrics, exploiting multi-factor authentication based on multiple biometric features.

\paragraph {Hill Climbing attack}

These attacks are carried out by submitting synthetic representations of the victim's biometrics iteratively, until successful recognition is obtained.
The employed data are modified at each step based on the results of previous attempts, expressed in terms of matching scores assumed to be known to the attacker, with the goal of improving the resulting matching output \cite{maiorana2014hill}. 
The system we propose is immune against such attacks owing to the use of the FCS, by which the verification of biometric features is performed in the encrypted domain, so that matching information is not returned to the attacker.

\section{Conclusion and future works}
\label{conclusion}

%Registration, enrollment, and authentication are the three stages of the system.

We proposed a novel framework for achieving decentralized biometric authentication through blockchain technology and fuzzy commitments.
In comparison to other blockchain-based biometric systems, our approach is efficient and allows achieving complete decentralization, meaning that the system can continue to run on the blockchain even if all the original users (i.e., the users who started it) leave the network.
We verified that the system has the necessary requirements for confidentiality, integrity and availability, and discussed  possible attacks on the proposed framework to assess its security. 

Our solution uses the FCS to gain privacy on a public blockchain like Ethereum.
Instead of a network of data centres or a distributed database, we use a public, permissionless blockchain, which is resilient to seizure and censorship and can survive even if all the original actors disappear.
The protocol's open and interoperable nature using public blockchains makes it easy for new players to adopt, %without joining any consortium.
while the system security relies on established technologies like ECCs and FCS. 
%Users' authentication performance is FCS-like with distributed network benefits.

Our biometric system framework is promising, but scalability and interoperability must be considered.  A future extended version of this paper will address these issues. We expect the need to use blockchain algorithms via smart contracts and associated costs to remain low, since all complex computations associated with FCS and ECCs are performed off-chain.

%Essentially, we only need to deploy a well-structured smart contract on a robust peer-to-peer network.

%The proposed framework could be improved in the future by addressing these issues and testing it in real-world scenarios. %The suggested framework lays the groundwork for future studies and developments in this area.

%Finally saying, In fact, the use of blockchain technology in biometric systems provides a decentralized, secure, and transparent solution to perform user authentication.

\bibliographystyle{IEEEtran}
\bibliography{Ref.bib}

\end{document}

%% file: FCS.tex
Let us provide a brief description of ECCs. For a given prime power $q$, an $(n,k)$ linear block code is a $k$-dimensional subspace of the  $n$-dimensional vector space  over the finite field $\mathbb{F}_q^n$. This implies that the code consists of $q^k$ distinct codewords of length $n$ over the field $\mathbb{F}_q$.
An encoder for an $(n,k)$ linear block code  maps any $q$-ary information vector of length $k$ into a $q$-ary codeword of length $n$. Given a certain received vector, the decoder tries to infer the original information vector, by performing error correction.

%%%%%%%BASICALLY THE SAME AS ALGORITHM 1

%\cite{juels1999fuzzy,chauhan2019improved}

FCSs rely on ECCs. In particular, the FCS we employ  \cite{juels1999fuzzy} is outlined  in Algorithm \ref{algoFCS}. Both the ECC (and the corresponding encoding and decoding functions) and the hashing function are assumed to be publicly available. In order to commit a codeword by using a vector $x$ of length $n$, we need to follow the procedure described next. First, we generate a witness $w$ of length $k$ (uniformly at random, for example), and encode it according to the chosen ECC, obtaining a codeword $c$ of length $n$. Then, the offset $\delta$ is computed by XORing the obtained codeword and $x$. The commit function returns the hash of the witness, and the offset.
 
 In order to decommit the codeword with a fuzzy version of $x$, denoted as $x'$, the offset and $x'$ are XORed, and the result is decoded. If the digest of the resulting information vector  equals the digest of the original witness, decommitment is deemed as successful.

\begin{algorithm}
\centering
\resizebox{0.4\textwidth}{!}{
\fbox{
\begin{minipage}{0.4\textwidth}
\DontPrintSemicolon
\SetKwProg{Fn}{Function}{:}{}
\SetKwFunction{GenerateRandomWitness}{GenerateRandomWitness}
\SetKwFunction{ECCEncode}{ECCEncode}
\SetKwFunction{ECCDecode}{ECCDecode}
\SetKwFunction{HashFunction}{Hash}
\SetKwFunction{Commit}{CommitFSC}
\SetKwFunction{Open}{OpenFSC}

\vspace{2mm}
\textbf{Public:} \HashFunction{}, \ECCEncode{}, \ECCDecode{}, \GenerateRandomWitness{}\;
\vspace{2mm}

\Fn{\Commit{$x$}}{
    $w \gets$ \GenerateRandomWitness{}\;
    $c \gets$ \ECCEncode{$w$} \quad \textcolor{gray}{\textit{$\#$ Encode witness}}\;
    $\delta \gets c \oplus x$  \textcolor{gray}{\textit{$\#$ Compute offset}}\;
    \KwRet (\HashFunction{$w$}, $\delta$) \textcolor{gray}{\textit{$\#$ Return hash of witness and offset}}\;
}\vspace{2mm}

\Fn{\Open{$\delta$, $x'$, \HashFunction{$w$}}}{
    $c' \gets x' \oplus \delta$ \textcolor{gray}{\textit{$\#$ Recompute offset with $x'$}}\;
    $w' \gets$ \ECCDecode{$c'$}\;
    \uIf{\HashFunction{$w'$} = \HashFunction{$w$}}{
        \KwRet True \textcolor{gray}{\textit{$\#$ Open successful}}\;
    }
    \Else{
        \KwRet False \textcolor{gray}{\textit{$\#$ Open failed}}\;
    }
}
\vspace{2mm}

\end{minipage}
}
}
 \vspace{1em}
 \caption{Fuzzy Commitment Scheme.}
 \label{algoFCS}
\end{algorithm}